\documentclass[aps,pra,amsmath,amssymb,reprint,letterpaper]{revtex4-1}

\usepackage{graphicx}
\usepackage{color}
\usepackage{natbib}
\setcitestyle{super}

\begin{document}

\title{Examination of Cluster Based Methods to Fermionic Systems}

\author{Jeffrey R. Keyes}
\affiliation{Department of Chemistry, Wesleyan University, Middletown,
  CT 06459}

\author{Carlos A. Jim\'enez-Hoyos}
\affiliation{Department of Chemistry, Wesleyan University, Middletown,
  CT 06459}

\begin{abstract}
  We present the cluster mean field(cMF) and its extensions into perturbation
  theory(cPT) and coupled cluster(cCC) methods for fermionic systems. We discuss
  how the commutation rules for fermionic systems modifies the method as well as
  other computational details. We then perform calculations on the Hubbard model
  and a few select conjugated molecules to demonstrate the efficacy of the
  method. We will also use these examples to compare cluster methods to CAS-SCF.

\end{abstract}

\maketitle

\section{Introduction}
Many systems of interest to chemists have stongly correlated electrons and
therefore require a multireference approach to adequately describe, these
include molecules with conjugation or metal clusters. The most common and
straight forward method is Complete Active Space Self Consistent
Field(CAS-SCF)\cite{zhao_casscfcaspt2_2009,sayfutyarova_constructing_2019,helmich-paris_benchmarks_2019}
where an active space is chosen to include all strongly correlated orbitals
often determined by chemical intuition, then all configurations that vary
occupation of these orbitals are used as a basis. For conjugated systems the
active space is typically the orbitals in the pi system and for metals clusters
are the d orbitals from each metal. A major drawback of this method is the
exponentially increase size of the Hilbert space as the size of the active space
increases, severely limiting the size of systems that can be studied with
CAS-SCF.

Several methods have been proposed to overcome this limitation such as selected
configuration interaction(sCI)\cite{levine_casscf_2020,smith_cheap_2017}, full
configuration interaction quantum monte
carlo(FCIQMC)\cite{weser_chemical_2021,liebermann_importance-sampling_2022,thomas_stochastic_2015},
and the density matrix renormalization
group(DMRG)\cite{sharma_density_2019,marti_density_2010,wouters_density_2014}
wavefunction. sCI methods define an active space similar to CAS-SCF, but then
only includes a subset of configurations according to some criterion, such as
the off diagonal values of the hamiltonian. sCI is limited by two things, the
hamiltonian is diagonalized each step which limits the number of configurations
that can be included and the criterion for determining which determinants are
used may exclude states necessary for an adequate description of the molecule.
FCIQMC is a method where the active space is sampled by a set of descrete signed
walkers that can cover a very large number of configurations and stochastically
evolve in imaginary time to project out excited states. The major drawback of
FCIQMC is the famous sign problem where the use of too few walkers results in
unphysical bosonic solutions, which may require a large number of walkers to
resolve. Lastly, the DMRG wavefunction factorizes the FCI wavefunction in a way
that takes advantage of the locality of correlations. The problem with DMRG is
that the method only has favorable scaling when the system under study is one
dimensional, which is obviously not always the case.

To remedy the active space problem, we propose using the recently developed
cluster meanfield(cMF) and the related cluster perturbation theory(cPT) and
cluster coupled cluster(cCC) methods.
\cite{jimenez-hoyos_cluster-based_2015,papastathopoulos-katsaros_coupled_2022,papastathopoulos-katsaros_symmetry-projected_2023}
The main idea of these methods is to break up the active space into a set of
smaller active spaces called clusters. These clusters should contain the
orbitals that are most correlated with each other, for example a conjugated pi
system could be broken into individual double bonds developing a set of 16
states where the bonding and anti-bonding states are filled in all possible ways
with anywhere from zero to four electrons. The cMF wavefunction is then taken as
the product state of cluster states optimized to minimize the energy. This
wavefunction interacts electrons within a cluster exactly while treating
interactions between clusters at the mean field level. If the clusters are
chosen well, then a significant portion of the correlation can be taken care of
at the mean field level while the size of the Hilbert space is much smaller
during the mean field optimization, then the remaining correlation is small
enough to be gathered with weak correlation methods like PT and CC.

This method was recently applied to the Heisenberg spin lattice in the
thermodynamic limit using up to cPT7 and cCCSDTQ5. We chose the Heisenberg spin
lattice since it is simpler to implement cluster methods in bosonic systems and
it is more strongly correlated than most molecular systems so it would act as a
good test of the capabilities of clusters methods. In that study, the energy
changes with increasing order of cPT and cCC steadily decreased in magnitude
indicating that both the PT and CC series were converging. Further, the energy
and site magnetization were in good agreement with values from other studies
that used high levels of theory. These results gave us confidence to proceed to
molecular systems.

In this paper, we will present the method and discuss important details of the
implementation, emphasizing the changes that need to be taken for fermionic
systems. The current implementation does not include rotations of active
orbitals with occupied or virtual orbitals, so the total active space will be
fixed for all calculations. This extension of the method will be addressed in
future work. We do allow rotations between orbitals belonging to different
clusters. We will then test the method on both the Hubbard model and some select
conjugated systems that will demonstrate the efficacy of the method.

\section{Theory}\label{sec:Theory}
\subsection{Cluster Mean Field}
The details of the formalism and optimization of cMF has been given
elsewhere\cite{jimenez-hoyos_cluster-based_2015}, so we will present only the
main ideas. The first step in applying the method is to partition the single
particle states of the active space into sets called clusters. The method will
be most effective if orbitals that are strongly correlated are placed together,
for example one could use localized orbitals and partition based on proximity.
In this paper, we assume that all orbitals are orthogonal and that clusters have
no overlap. We then create a set of states, called cluster states, which are
linear combinations of all possible slater determinants formed using orbitals
from the cluster including anywhere from zero electrons to enough electrons to
fill all orbitals. Including cluster states of different particle number is done
to allow for charge transfer effects. To build the total wavefunction, we define
a set of cluster operators, $A^{\dagger}_{p,c}$($A_{p,c}$) that
create(annihilate) the pth cluster state from cluster c. These states can be
ordered by ascending energy such that $A_{0,c}$ is the annihilation operator for
the ground state of cluster c. The cMF wavefunction is then given by:
\begin{equation}
  |\Phi_0\rangle = \prod_{i=1}^{N}A_{0,i}^\dagger |-\rangle
\end{equation}
where $N$ is the number of clusters and $|-\rangle$ is a theoretical vacuum
where no clusters are occupied. These states are then chosen to minimize the
energy, $\langle\Phi_0|H|\Phi_0\rangle$. Note that the physical vacuum is a
state formed as the product of cluster states that contain no electrons and
should not be confused with the theoretical vacuum. We will discuss the
commutation of these operators below. In this paper, we will also restrict each
cluster state to only use determinants that have the same number of electrons,
which will be convenient when calculating the correlation energy as we will
discuss below.

Using this formalism we can build the full Hilbert space by exciting clusters
out of the reference state, $|\Phi_0\rangle$. For example, we can produce a 
singly excited state as:
\begin{equation}
  |\Phi_c^a\rangle = A^\dagger_{a,c}A_{0,c}|\Phi_0\rangle
\end{equation}
This can then be extended to higher excitations to build the full Hibert space.
All physical states are those that have all clusters occupied by some cluster
state, so the cluster operators only have meaning when carried in pairs. 

\subsection{Cluster States and Operators}
Since the number of electrons in each cluster state can be different, the
commutation rules are complicated and need to be handled with care. To write
operators using the cluster operators, we need to map between the fermionic
operators and cluster operators. We will not define the precise relationships
here beyond to say that fermionic operators connect clusters states that include
determinants that differ by the occupation of one orbital. This means we can
represent the annihilation operator of the pth orbital in the wth cluster as:

\begin{equation}
  a_{p,w} = \sum_{nm}C_{mn}^w A^\dagger_{m,w}A_{n,w}
\end{equation}

where $n,m$ runs over cluster states and $C_{mn}^w =
\langle\Phi_w^m|a_{p,w}|\Phi_w^n\rangle$ are expansion coefficients that account
for both the details of the cluster states but also the sign required to be
consistent with the order that the orbitals were filled. A similar relation can
be built for creation operators. If we were to represent a two electron operator
in the cluster representation, it would include up to four cluster terms.
However, if the cMF accounts for enough of the correlation, then the three and
four cluster terms should have small contributions to the cPT and cCC energy. In
this paper, when doing cPT or cCC we limit the hamiltonian to include two
cluster terms only, so we will represent the two electron integrals as:
\begin{equation}
  V = \sum_{wxpqrs} V_{pqrs}^{wx}A^\dagger_{p,w}A_{r,w}A^\dagger_{q,x}A_{s,x}
\end{equation}
where $w,x$ run over clusters and $p,q,r,s$ run over cluster states. Since the
two electron integrals respect particle number symmetry, each term of the sum
can be classified into one of three cases, those where two electrons are
transferred between clusters, those that transfer one electron between clusters,
and those that transfer no electrons. In the case where two electrons are
transferred, the operators $A^\dagger_{p,w}A_{r,w}$ represent either two
fermionic creation operators or two fermionic annihilation operators, therefore
this pair of cluster operators will commute with all other cluster operators on
different clusters. For the case when one electron is transferred,
$A^\dagger_{p,w}A_{r,w}$ must represent either one or three fermionic operators,
so it will anticommute with other pairs of cluster operators that change the
number of electrons by one. Finally for the case where no electrons are
transferred, $A^\dagger_{p,w}A_{r,w}$ must represent two fermionic operators so
it will always commute. This motivates a separation of the integrals:
\begin{equation}
  \label{eq:decomp}
  V_{pqrs}^{wx} = \bar{V}_{pqrs}^{wx} + \bar{W}_{pqrs}^{wx}
\end{equation}
for the commuting and anticommuting terms with the symmetry:
\begin{equation}
  \bar{V}_{pqrs}^{wx} = \bar{V}_{qpsr}^{xw}
\end{equation}
\begin{equation}
  \bar{W}_{pqrs}^{wx} = -\bar{W}_{qpsr}^{xw}
\end{equation}
To illustrate how to account for these commutation rules when evaluating
expectation values consider the following contraction which contributes to
$\langle\Phi_0| V^3|\Phi_0\rangle$ = $\langle V^3\rangle$:
\begin{multline}
  \sum_{wxypqs}V_{00pq}^{wx}V_{p00s}^{wy}V_{sq00}^{yx}
  \\ \langle A^\dagger_{0,w}A_{p,w}A^\dagger_{0,x}A_{q,x}
          A^\dagger_{p,w}A_{0,w}\\
          A^\dagger_{0,y}A_{s,y}
          A^\dagger_{s,y}A_{0,y}A^\dagger_{q,x}A_{0,x} \rangle
\end{multline}
The zeros are the reference state for each cluster. To evaluate this correctly,
we insert Eq. \ref{eq:decomp} for each integral and distribute. The sign of each
term is then found by grouping the operators for each cluster to yield:
\begin{equation}
  \langle A^\dagger_{0,w}A_{p,w}A^\dagger_{p,w}A_{0,w}
          A^\dagger_{0,x}A_{q,x}A^\dagger_{q,x}A_{0,x}
          A^\dagger_{0,y}A_{s,y}A^\dagger_{s,y}A_{0,y} \rangle
\end{equation}
accounting for signs depending on which combination of symmetric and
antisymmetric integrals are being used. In this case, the expectation is:
\begin{equation}
  \sum_{wxypqs}\bar{V}_{00pq}^{wx}\bar{V}_{p00s}^{wy}\bar{V}_{sq00}^{yx}
             - \bar{W}_{00pq}^{wx}\bar{W}_{p00s}^{wy}\bar{W}_{sq00}^{yx}
\end{equation}
All other combinations produce a zero.

\subsection{Perturbation Theory and Coupled Cluster}
Perturbation theory and coupled cluster do not fundamentally change in the
cluster formalism. For perturbation theory, the unperturbed hamiltonian 
which is diagonalized by the cMF optimization is:
\begin{multline}
  H_0 = \sum_w \bigg[\sum_{pr}h_{pr}a_{p,w}^\dagger a_{r,w}\\
      + \delta_{wx}\sum_{pqrs} \langle pq|rs\rangle a_{p,w}^\dagger a_{q,x}^\dagger a_{s,x} a_{r,w}\\
      +\sum_{x\neq w}\rho_{qs}^x(\langle pq|rs\rangle - \langle pq|sr\rangle)a_{p,w}^\dagger a_{r,w} \bigg]
\end{multline}
where $\rho$ is the one particle density and $h$ is the one particle part of the
hamiltonian. Notice how the interaction of electrons within a cluster is exact,
but those between clusters are kept at a mean field level. This then defines the
perturbation by $V = H - H_0$. We found that it was more efficient to calculate
all expectation values directly using projections with the energy evaluation:
\begin{equation}
  E^{(n)} = \langle\Phi_0|V\left[\frac{Q}{E_0-H0}V\right]^{n-1}|\Phi_0\rangle + \cdots
\end{equation}
where
\begin{equation}
  Q = \sum_{wa}|\Phi_w^a\rangle\langle\Phi_w^a| + \sum_{wxab}|\Phi_{wx}^{ab}\rangle\langle\Phi_{wx}^{ab}| + \cdots
\end{equation}
rather that follow a diagrammatic approach. The particular expectation values
needed for each order can be gathered from the bracketing procedure outline by
Shavitt and Bartlett\cite{shavitt_many-body_2009}.

For CC, the only additional change worth mentioning is that due to our
restriction to two cluster operators, any terms that include more than two
excitation operators will be zero since either the two electron integral will be
zero or the term will be disconnected.

\subsection{Systems and Tiling}
\begin{figure}[h]
  \includegraphics[width=0.45\textwidth]{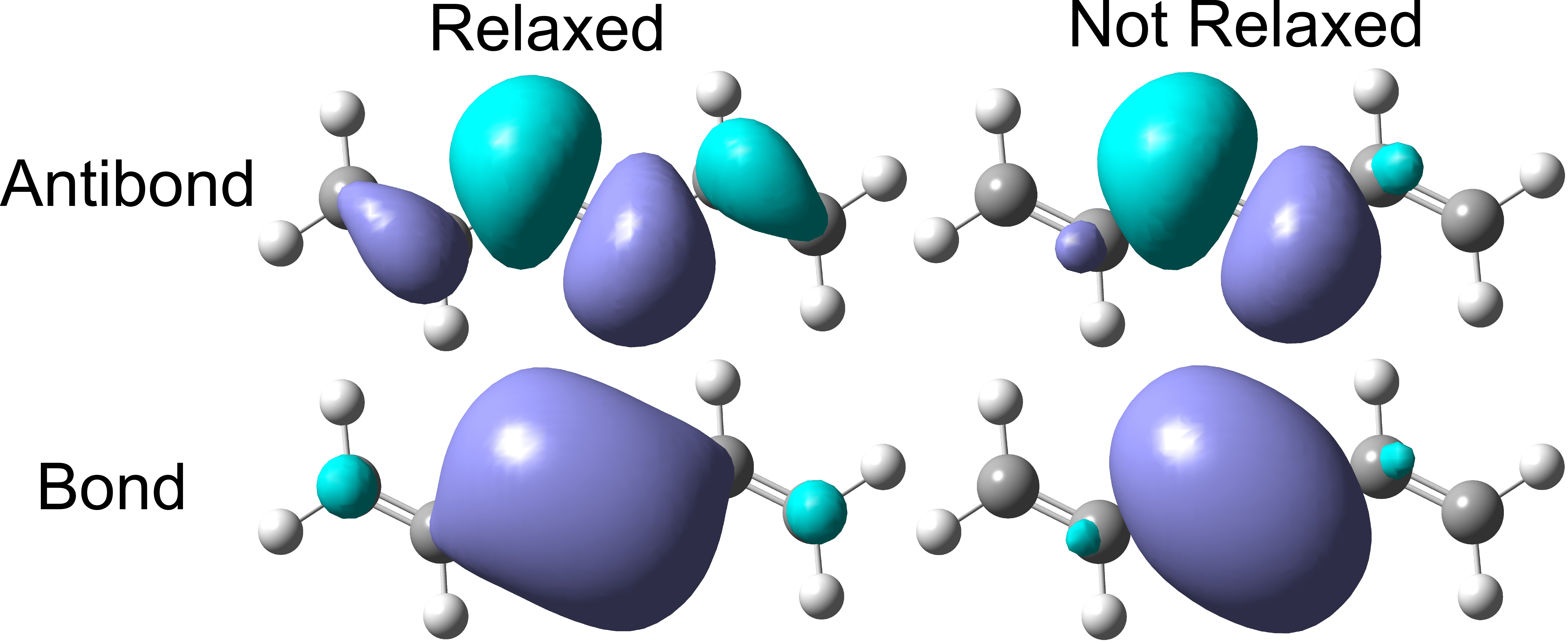}
  \caption{Orbitals used for central cluster of hexatriene before and after
  optimization. }
  \label{fig:orb}
\end{figure}

The cMF formalism will work best on systems where the active space can be a
priori partitioned into sets of orbitals that are more correlated within the set
than between sets. The first type of system that we will first examine is
conjugated systems where each cluster will contain the bonding and anti-bonding
orbitals from each double bond. This will produce clusters that each have 16
states. For the smaller molecules with less than 10 carbons, we will use the
CAS-SCF active space rotated to resemble bonding and anti-bonding orbitals to
form the clusters. This is accomplished by producing bonding and anti-bonding
orbitals using the least diffuse pz atomic orbitals from the two carbons, these
are then orthonormalized and projected into the CAS active space. An example is
shown in Fig. \ref{fig:orb} for the central cluster of hexatriene. The typical
behavior for polyenes is to mix the bonding orbital of a cluster with the
anti-bonding orbital of the neighboring cluster and vice versa. For larger
systems, we perform the same procedure using UHF natural orbitals with
occupations that differ significantly from 0 or 2. As mentioned above, we
restrict each cluster state to use configurations with the same number of
electrons which makes the sign determination of each term easier.

In a similar light, we will perform calculations on the one dimensional Hubbard
hamiltonian with 30 sites. The clusters will consist of the on site orbitals for
each neighboring pair, producing clusters with 16 states, similar to the
conjugated molecules.

For the cMF calculations, we will present results with both relaxed and
unrelaxed orbitals. Here, the unrelaxed orbitals are those produced by the above
mentioned procedure with no further alteration. The relaxed orbitals are those
where we allow for rotations of orbitals between clusters. An example can be
seen in Fig. \ref{fig:orb}, these relaxed orbitals tend to delocalize into the
neighboring clusters. The current implementation does not include rotations
between the active orbitals and the occupied or virtual orbitals. 

\subsection{Scaling}
For cPT, the computational scaling of each diagram is similar to the usual MP
diagrams with a few changes. First, the number of ``occupied" states is always
1, since for each cluster there is only one occupied cluster state. Instead, the
diagram has to be evaluated for every set of viable clusters, which tends to be
less than the number of occupied labels that a diagram has. The particular
combinations can be found using basic graph theoretical techniques. Lastly,
since the integrals are being split into commuting and anti-commuting terms,
each cPT$n$ diagram may have to evaluated up to $2^n$ times to account for all
combinations, though most diagrams use fewer. The situation for cCC is similar
with the exception that every term has only one integral, so each term may need
to be evaluated twice.

The storage scaling of this method is dominated by the two cluster integrals,
for $n$ cluster states and $m$ clusters, there will be $n^4m^2$ integrals. 

\section{Computational Details}
Calculations of cMF, cPT, and cCC energies were done using in-house code,
geometry optimizations were done in the Gaussian 16 program, and CAS-SCF
calculations were done with the pySCF software. Geometries for all molecules
were computed with the cc-pvdz basis and unrestricted density functional
theory(DFT) with the B3LYP functional. Linear polyenes all have C2h point group
symmetry and the polycyclic aromatic hydrocarbons(PAH) have D6h symmetry.

\section{Results and Discussion}\label{sec:Results} 

  

\begin{table*}[t]
  \caption{ Error in Energy(mHa) per carbon compared to 2cCASCI for cluster
  methods with and without relaxing the cluster states. All calculations
  performed using CAS-SCF active space. \label{tab:small}}
  \begin{ruledtabular}
  \begin{tabular}{l r|r r r r r r}
  
  Molecule  & 4cCASCI & cMF & cPT2 & cPT3 & cPT4 & cPT5 & cCCSD\\
  \hline
  Not Relaxed &&&&&&& \\
  butadiene  & 0.0 & 8.85508047 & 3.3104363 &  1.48952783 & 0.63595067 & 0.29831711 &-0.70256333\\
  hexatriene &-0.58671993 &11.86203567 & 4.56811131&  2.12776015&  0.85597072  &0.30677783 &-1.24976842\\
  benzene    &0.6647978& 33.45075728& 21.23460101 &14.76778668 & 9.76493558 & 6.33437177 &-3.742064\\
  octatetraene    &-1.03402714 & 13.41429537 & 5.27391804 & 2.54071264 & 1.07242184 & 0.42219022& -1.43201131\\
  \hline
  Relaxed &&&&&&& \\
  butadiene  & 0.0 & 1.2396099  & 0.46315603 & 0.25590696 & 0.14837084 & 0.09586104 & 0.02440985 \\
  hexatriene &1.69733563 & 3.45884066  &0.87280364 & 0.39127274  &0.15570659  &0.07912682 & 0.02671819\\
  benzene    &-1.71268291& 3.86673396 & 0.45113046 & 0.7077482 &  0.23865598 & 0.2787179 &  0.15163605 \\
  octatetraene    &2.90373308 &4.9856245  & 1.37000972 & 0.60095333 & 0.20545836 & 0.08706456 &-0.05601636\\
  \end{tabular}
  \end{ruledtabular}
\end{table*}

\begin{figure}[h]
  \includegraphics[width=\columnwidth]{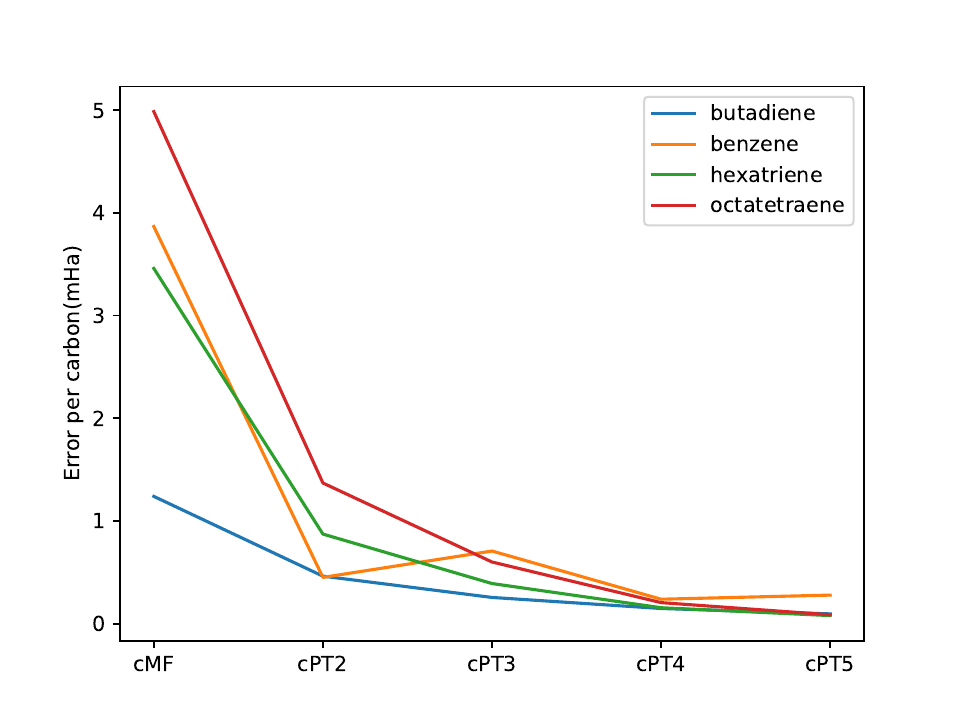}
  \caption{Error of cPT calculations for small molecules using relaxed orbitals.}
  \label{fig:small}
\end{figure}

We begin by comparing cluster methods to CAS calculations on small conjugated
molecules. In Table \ref{tab:small}, we show the error per carbon for up to cPT5
and cCCSD both with and without relaxation of the clusters states by
intercluster rotations of the orbitals. The errors presented are given with
respect to a CASCI calculation done with the CAS-SCF orbitals but with the three
and four cluster terms removed, which we refer to as 2cCASCI. As can be seen,
the restriction to two cluster interactions introduces a small error compared to
the 4cCASCI, which includes the three and four cluster terms, when there is more
that two double bonds. The error becomes larger when relaxing the orbitals
because the relaxation will delocalize the cluster states and thereby increase
coupling. This error could be removed by including the three and four cluster
interaction in the cPT calculation, however this would increase the required
computational effort. In our tests, the three and four cluster terms had a
negligible effect on the cPT2 energy, but we expect that it will become more
significant for cPT3 and higher. We will leave it to future investigations to
determine whether this error has significant effect on properties and should be
corrected.

We have plotted the errors for the relaxed calculation in Fig. \ref{fig:small}.
Comparing the three linear polyenes, the errors increase with system size, which
may be the effect of having more pairs of clusters to interact, but the larger
molecules error is converging faster such that the differences in errors of
higher order cPT and cCCSD calculations are very small, suggesting that the
method is extensive. Comparing the relaxed and unrelaxed calculations, it is
clear that relaxing the orbitals is important. The error per carbon at the cMF
level is significantly increased, the cPT series converges much slower, and the
cCCSD value is further off without relaxation. This is consistent with previous
investigations of cluster methods. The error seems to be more significant in
benzene than the n-polyenes, this is likely due to the fact that benzene has two
equivalent resonance structures, so the partitioning of the orbitals into double
bonds is less appropriate for benzene. Note that while the PT series will
converge slower, we see no reason the converged value would be worse than for
systems with one resonance structure.

\begin{table*}[t]
  \caption{ Error in Energy(mHa) per carbon compared to CAS-SCF for cluster
  methods with single orbital clusters(details in text). \label{tab:single}}
  \begin{ruledtabular}
  \begin{tabular}{l r r r r }
  
  Molecule  & HF & MP2 & MP3 & MP4 \\
  \hline
  butadiene  & 14.18082988 & 6.81568833 & 3.19000589& 1.42301511\\
  hexatriene &14.12973757 & 6.75202132 & 3.20284331 & 1.46998662\\
  benzene    &12.39057563 & 5.22830043 & 2.6332501 &  1.41064235\\
  octatetraene &14.11414573 & 6.70952084 & 3.20932507 & 1.50138325\\
  \end{tabular}
  \end{ruledtabular}
\end{table*}

The next question we want to address is whether the grouping of the bonding and
anti-bonding orbitals into a cluster provides substantial benefit. To this end,
we performed HF and standard M\o ller-Plesset perturbation theory calculations
on the small polyenes shown in Table \ref{tab:single}. To make the values
comparable to the cluster calculations, we correlated only the orbitals in the
active space, freezing the other occupied orbitals and virtual orbitals. This
calculation does include the three and four cluster terms, so the comparison is
only approximate. Comparing with the values in Table \ref{tab:small}, the errors
are approximately 10 times bigger than those using the cluster methods. This
demonstrates that the clustering of orbital pairs has a substantial effect of
the quality of the wavefunction at the cMF level, setting up a faster converging
cPT series. This effect could be extended by including more double bonds per
cluster, but this will increase the number of cluster states from 16 to 256
which will increase the computational requirements for both the mean field and
correlated calculations. We will discuss a situation where extending the
clusters is needed below.

  

\begin{table*}[t]
  \caption{Correlation energy(Ha) for cluster methods with relaxed cluster
  states. RHF values are -462.511402, -616.299771, -770.087943, -916.009941, and
  -2055.867568 Ha in the order shown below. \label{tab:large}}
  \begin{ruledtabular}
  \begin{tabular}{l r r r r r r}
  
  Molecule  & cMF & cPT2 & cPT3 & cPT4 & cPT5 & Lit.\\
  \hline
  C$_{12}$H$_{14}$  & -0.129355 & -0.184393 & -0.197937 & -0.205855 & -0.208579  & -0.149858\cite{ghosh_orbital_2008}\\
  C$_{16}$H$_{18}$  & -0.169271 & -0.242770 & -0.261678 & -0.273164 & -0.277335  & -0.199491\cite{ghosh_orbital_2008}\\
  C$_{20}$H$_{22}$  & -0.209043 & -0.297650 & -0.320762 & -0.335200 & -0.340544  & -0.249169\cite{ghosh_orbital_2008}\\
  C$_{44}$H$_{46}$  & -0.446728 & -0.600854 & -0.639890 & -0.667177 & -0.677458  & -\\
  Coronene(1)       & -0.119870 & -1.024297 & -1.335185 & -1.564843 & -1.713366  & -0.27\cite{thomas_stochastic_2015}      \\
  Coronene(2)       & -0.154659 & -1.410654 & -1.739200 & -2.177562 & -2.490637  & -0.27\cite{thomas_stochastic_2015}      \\
  Circumcoronene    & -0.289304 & -5.337128 & -7.485396 & -10.515621& -13.405619 & -      \\
  \end{tabular}
  \end{ruledtabular}
\end{table*}

\begin{figure}[h]
  \includegraphics[scale=0.25]{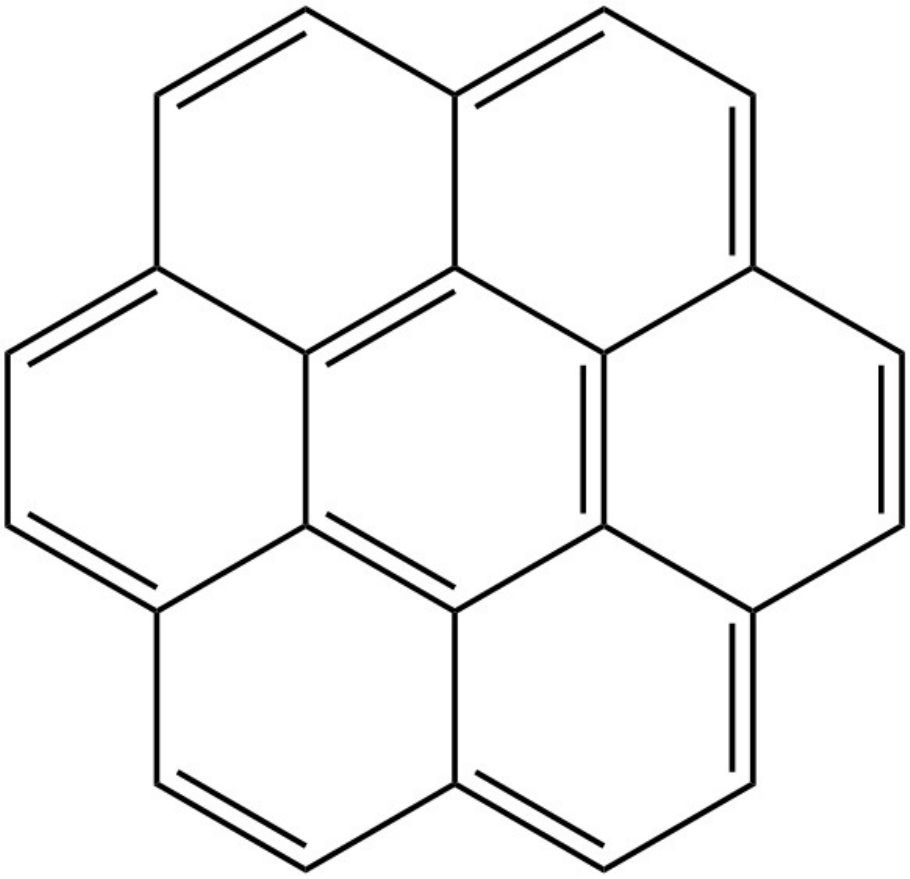}
  \includegraphics[scale=0.25]{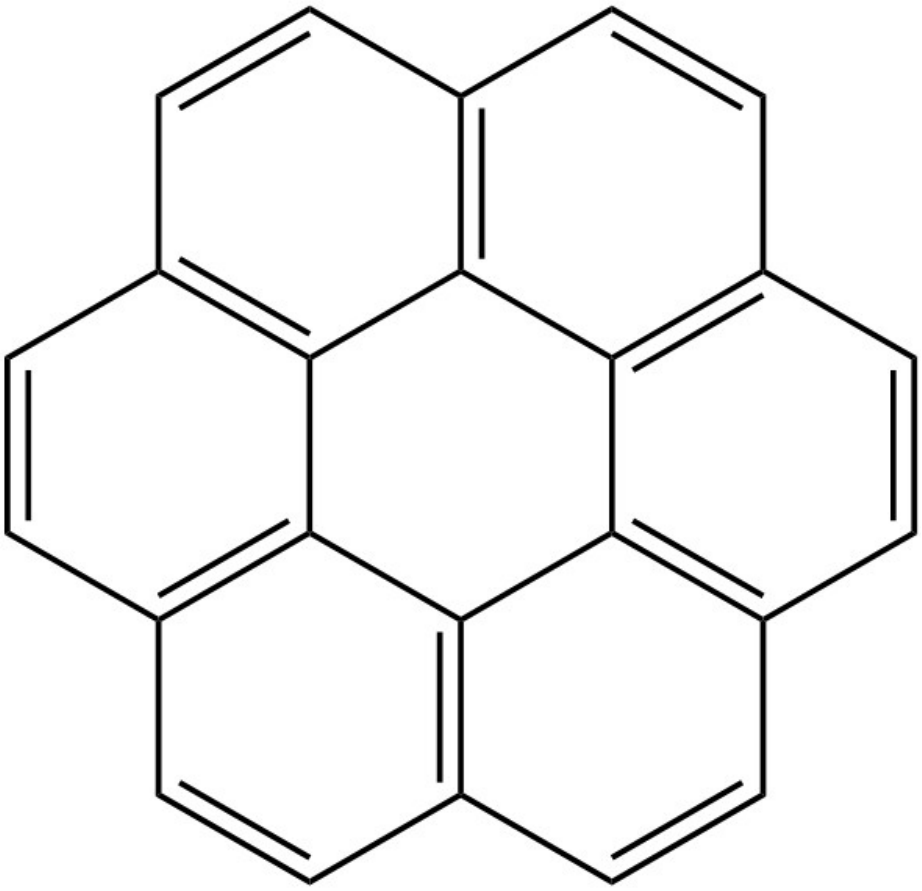}
  \includegraphics[scale=0.5]{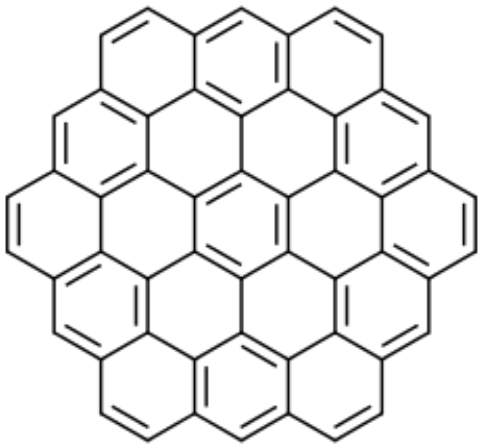}
  \caption{Lewis structure coronene(1), coronene(2) and circumcoronene}
  \label{fig:PAH}
\end{figure}

We now look at larger conjugated systems, Table \ref{tab:large} shows the
cluster results for linear polyenes up to C$_{44}$H$_{46}$ and two larger
polycyclic aromatic hydrocarbons(PAH), coronene(C$_{24}$H$_{12}$) and
circumcoronene(C$_{54}$H$_{18}$), shown in Figure \ref{fig:PAH}. Here, we see
similar results for the polyenes to the smaller systems discussed above, namely
the PT series for all systems are converging nicely. The 2cCASCI results for
C$_{12}$H$_{14}$ and C$_{16}$H$_{18}$ are -0.210892 and -0.281497 respectively,
so the PT series is approaching the correct limit. We were unable to converge
the cCCSD equations for the larger molecules, in the case of the linear
polyenes, the energy of the iterations were oscillating around reasonable
values. For example, the C$_{12}$H$_{14}$ correlation energies were around
-0.2133 Ha and the C$_{16}$H$_{18}$ energies were around -0.2863 Ha, both are
just below the 2cCASCI which is consistent with the smaller polyenes. We are not
sure why the cCCSD equations are not converging and will leave it to future
investigations. We also include DMRG results from the
literature\cite{ghosh_orbital_2008}, there may be small difference in the
molecular geometry and the orbitals used, so comparison should be done with
caution. In all three cases the cMF captures about 85\% of the correlation
energy and the cPT5 energy difference is about 4.5 mHa per carbon, again
suggesting that the error due to restriction to two cluster terms is linear with
system size. One important advantage that cluster methods have over DMRG is that
the computational cost does not increase as the system dimensionality increases
whereas the bond dimension in DMRG calculations needs to be increased to
maintain accuracy for high dimensional materials.

We now turn our attention to the PAH systems. One difference between benzene and
coronene is that the two resonance structures of benzene are related by symmetry
where coronene has multiple resonance structures that are not related by
symmetry. We ran cluster calculations starting from the two resonance structures
of coronene show in Fig. \ref{fig:PAH} and found that the cMF energy differed by
about 35 mHa, indicating that the choice of clusters can have a significant
impact on the result. An FCIQMC-SCF calculation\cite{thomas_stochastic_2015} of
coronene has been done where they calculated the total energy to be about
-916.28 Ha. Since their calculation included an optimization of the active space
the comparison is only approximate, but the cMF values that we have are about
half of the FCIQMC values, so the cMF wavefunction includes a good portion of
the correlations. 

\begin{figure}[h]
  \includegraphics[scale=0.9]{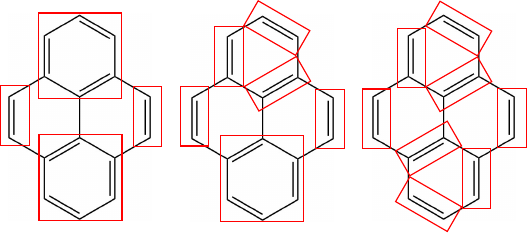}
  \caption{Cluster choice for pyrene when using 4(left), 6(center), or 8(right) clusters}
  \label{fig:pyrene}
\end{figure}

However, looking at the coronene cPT values it is clear that they are
unreasonably large indicating that the cPT will diverge, we also found that the
cCCSD energies diverged. The case for circumcoronene is very similar, producing
a reasonable cMF value and unphysically large cPT values. When examining the
contributions to the PT2 energy, we find that the terms involving a transfer of
an electron between clusters are the ones that are diverging. The cause of this
is our restriction to two cluster interactions only, when transferring an
electron there will be repulsion from the electrons in the two clusters involved
in the transfer, which is captured in the two cluster terms. However, there is
also the repulsion from the mean field of the rest of the clusters that is not
included in the above results. To demonstrate that including these terms
corrects the divergence, we performed cPT2 calculations using the full
hamiltonian on pyrene, coronene and circumcoronene. We consider here several
cluster choices where some benzene rings are gathered into a single cluster. For
example in pyrene we considered the three choices shown in Fig. \ref{fig:pyrene}
and label them by the number of clusters in Table \ref{tab:pyrene}. We can do a
similar partitioning for the coronene(2) structure(Fig. \ref{fig:PAH}) where the
three benzene rings on the exterior can be chosen as a cluster, collected in
Table. \ref{tab:coro}. We also did a calculation on circumcoronene using the
clusters as above, we find that the cMF gathers -0.675401 Ha of correlation
energy while cPT2 gathers -0.872290 Ha of correlation energy. Comparing these
calculations with those above, it is clear that by incorporating the three
cluster terms produces much more reasonable correlation energies. In fact, our
coronene calculations are recovering about 90\% of the correlation energy
compared with the FCIQMC value from the literature.\cite{thomas_stochastic_2015}
We are currently working on an approach to incorporate the three and four
cluster terms without substantial increases in cost and will leave the topic to
future publications.

Before comparing cluster choices, we want to provide a comparison of the
computational cost of our cPT5 and the FCIQMC\cite{thomas_stochastic_2015} from
the literature using the same size active space. As reported, each step of their
calculation required calculating the FCIQMC energy and the evaluating the
orbital gradient, each step took about 3840 CPUh and the full optimization took
about 40000 CPUh using about 3 GB of RAM. Our cPT5 calculation required about
0.1 GB to store the $16^412^2$ integrals and approximately 50 CPUh. Our current
implementation of cPT5 is not capable of performing the calculation with the
double bonds of each ring collected, so we cannot provided that comparison.
While we cannot directly compare the CPUs used or the relative efficiency of the
two implementations, it is clear that our approach is considerably cheaper.
Unfortunately, we cannot provide the same comparison with the
DMRG\cite{ghosh_orbital_2008} calculations since timings were not reported.

Comparing the cluster choice for pyrene and coronene, collecting the pi bonds of
each benzene ring has a substantial effect on the quality of the cMF
wavefunction. The cMF value for pyrene using 4 clusters is about 50 mHa below
that with 8 clusters and similar improvements are seen for coronene. However,
these gains are far less significant for the cPT2 values, gaining about 15 mHa
for pyrene and only 10 mHa for coronene. We expect the difference to continue
diminishing as higher orders are included. Considering that using larger
clusters can increase the cost of correlated calculations, it remains to be seen
what the optimal cluster choice is.

\begin{table}[h]
  \caption{Error in Energy(mHa) compared to CASCI for pyrene using different
  number of clusters. Perturbation energy computed with four cluster terms.
  RHF value is -611.808932 Ha \label{tab:pyrene}}
  \begin{ruledtabular}
  \begin{tabular}{l r r }
  
  Number of clusters  & cMF & cPT2 \\
  \hline
  4  & -0.154306 &  -0.180930 \\
  6  & -0.128521 & -0.172403  \\
  8  & -0.103509 & -0.163991 \\
  \end{tabular}
  \end{ruledtabular}
\end{table}

\begin{table}[h]
  \caption{Correlation Energy(Ha) of coronene using different number of
  clusters. Perturbation energy computed with four cluster terms. RHF value for coronene 
  is -916.009941
  \label{tab:coro}}
  \begin{ruledtabular}
  \begin{tabular}{l r r  }
  
  Number of clusters  & cMF & cPT2 \\
  \hline
  6  & -0.206727 & -0.257707 \\
  8  & -0.188811 & -0.256682 \\
  10 & -0.171471 & -0.251913 \\
  12 & -0.154661 & -0.247149 \\
  \end{tabular}
  \end{ruledtabular}
\end{table}

  

\begin{table}[h]
  \caption{ Error in Energy(T) per site compared to FCI for cluster methods
  with and without relaxing the cluster states. x indicates that the 
  calculation did not converge. \label{tab:hub}}
  \begin{ruledtabular}
  \footnotesize
  \begin{tabular}{l r r r r r r}
  
  U/T & cMF & cPT2 & cPT3 & cPT4 & cPT5 & cCCSD\\
  \hline
  Not Relaxed\\
  2   & 0.22808 & 0.03522 & 0.03522 & 0.00685 & 0.00685 & x \\
  4   & 0.16024 & 0.03712 & 0.03712 & 0.00810  & 0.00810  & x \\
  6   & 0.11783 & 0.03316 & 0.03316 & 0.00883 & 0.00883 & x \\
  8   & 0.09189 & 0.02834 & 0.02834 & 0.00819 & 0.00819 & x \\
  12  & 0.06310 & 0.02099 & 0.02099 & 0.00643 & 0.00643 & x \\
  \hline
  Relaxed\\
  2   & 0.02074 & 0.01302  & 0.01225 & 0.01041  & 0.00994 & 0.01048 \\
  4   & 0.05620  & 0.02556 & 0.02048  & 0.01319 & 0.01205 & 0.01260 \\
  6   & 0.06947 & 0.02813 & 0.01938 & 0.01296 & 0.01005 & 0.00726 \\
  8   & 0.06745  & 0.02629 & 0.01825 & 0.01132 & 0.00865 & 0.00335 \\
  12  & 0.05481 & 0.02058 & 0.01628 & 0.00783 & 0.00672 & 0.00033 \\
  \end{tabular}
  \end{ruledtabular}
\end{table}

The last system we will examine here is the one dimensional hubbard model with
30 sites, we present energy errors compared to the FCI results for U/T values
from 2 to 12 in Table \ref{tab:hub}. Similar to the polyenes, relaxing the
cluster states has a significant impact on the quality of the cMF energy and
accelerates the convergence of the cPT series. An important note when viewing
the results of Table \ref{tab:hub}, due to the finite range of the hubbard
potential the two cluster approximation we make is exact for the unrelaxed on
site basis, but the relaxed orbitals will introduce three and four cluster terms
so the cPT series will converge to a different value. Further, the hubbard
potential also makes the odd powers of the cPT series with the on site basis not
contribute which is why the cPT4 and cPT5 values are the same. Comparing results
for the different U/T values, the cMF wavefunction is more effective for higher
U/T values when using the onsite basis, which is to be expected since in the
high U/T limit, the solution isolates each particle onto one site. This
difference disappears for the cPT energies as the error flattens moving through
the cPT series. We also found that the cCCSD equations did not converge when
using the on site basis. When orbital relaxation is included, there is no
difference in the error of the cMF or cPT wavefunctions. Lastly, the cCCSD
equations did converge when relaxed orbitals were used, perhaps reflecting the
smaller amount of correlation left after the cMF optimization.

\section{Conclusion}
In this paper, we have adapted the cluster mean field wavefunction and its
extensions in perturbation theory and coupled cluster to fermionic systems. We
presented the theory and discussed details of implementation. We then applied
the method to several conjugated molecules and the hubbard model where we
demonstrated that the method provides a reasonable approximation of the CAS-SCF
value and all errors scale linearly, so the efficacy remains for large systems.
This data suggests to us that cluster methods may be a suitable low cost method
for strongly correlated systems with large active spaces.

Some directions that we are interested in for future investigations include
determining if the three and four cluster terms affect molecular properties
significantly or if they can be left out. We would also like to better
understand why the cCCSD equations are not converging for some systems and how
the choice of clusters affects these results. We are interested in extending the
theory to allow for rotations between the cluster states and the core and
virtual spaces. This could be done by simply using the core and virtual orbitals
as their own clusters, but this would significantly increase the cost, so
limiting the number of electrons transferred in or out of these spaces would be
necessary to limit the number of cluster states. 

\bibliographystyle{apsrev4-1}
\bibliography{cMF_mol}

@article{ghosh_orbital_2008,
	title = {Orbital optimization in the density matrix renormalization group, with applications to polyenes and beta-carotene},
	volume = {128},
	issn = {0021-9606},
	url = {https://doi.org/10.1063/1.2883976},
	doi = {10.1063/1.2883976},
	number = {14},
	urldate = {2024-02-06},
	journal = {The Journal of Chemical Physics},
	author = {Ghosh, Debashree and Hachmann, Johannes and Yanai, Takeshi and Chan, Garnet Kin-Lic},
	year = {2008},
	pages = {144117},
}

@article{thomas_stochastic_2015,
	title = {Stochastic {Multiconfigurational} {Self}-{Consistent} {Field} {Theory}},
	volume = {11},
	issn = {1549-9618},
	url = {https://doi.org/10.1021/acs.jctc.5b00917},
	doi = {10.1021/acs.jctc.5b00917},
	number = {11},
	urldate = {2024-02-06},
	journal = {Journal of Chemical Theory and Computation},
	author = {Thomas, Robert E. and Sun, Qiming and Alavi, Ali and Booth, George H.},
	year = {2015},
	note = {Publisher: American Chemical Society},
	pages = {5316--5325},
}

@article{papastathopoulos-katsaros_coupled_2022,
	title = {Coupled {Cluster} and {Perturbation} {Theories} {Based} on a {Cluster} {Mean}-{Field} {Reference} {Applied} to {Strongly} {Correlated} {Spin} {Systems}},
	volume = {18},
	issn = {1549-9618},
	url = {https://doi.org/10.1021/acs.jctc.2c00338},
	doi = {10.1021/acs.jctc.2c00338},
	number = {7},
	urldate = {2024-02-06},
	journal = {Journal of Chemical Theory and Computation},
	author = {Papastathopoulos-Katsaros, Athanasios and Jiménez-Hoyos, Carlos A. and Henderson, Thomas M. and Scuseria, Gustavo E.},
	year = {2022},
	note = {Publisher: American Chemical Society},
	pages = {4293--4303},
}

@article{jimenez-hoyos_cluster-based_2015,
	title = {Cluster-based mean-field and perturbative description of strongly correlated fermion systems: {Application} to the one- and two-dimensional {Hubbard} model},
	volume = {92},
	shorttitle = {Cluster-based mean-field and perturbative description of strongly correlated fermion systems},
	url = {https://link.aps.org/doi/10.1103/PhysRevB.92.085101},
	doi = {10.1103/PhysRevB.92.085101},
	number = {8},
	urldate = {2024-02-06},
	journal = {Physical Review B},
	author = {Jiménez-Hoyos, Carlos A. and Scuseria, Gustavo E.},
	year = {2015},
	note = {Publisher: American Physical Society},
	pages = {085101},
}

@article{papastathopoulos-katsaros_symmetry-projected_2023,
	title = {Symmetry-projected cluster mean-field theory applied to spin systems},
	volume = {159},
	issn = {0021-9606},
	url = {https://doi.org/10.1063/5.0155765},
	doi = {10.1063/5.0155765},
	number = {8},
	urldate = {2024-02-06},
	journal = {The Journal of Chemical Physics},
	author = {Papastathopoulos-Katsaros, Athanasios and Henderson, Thomas M. and Scuseria, Gustavo E.},
	year = {2023},
	pages = {084107},
}

@article{sayfutyarova_constructing_2019,
	title = {Constructing {Molecular} pi-{Orbital} {Active} {Spaces} for {Multireference} {Calculations} of {Conjugated} {Systems}},
	volume = {15},
	issn = {1549-9618},
	url = {https://doi.org/10.1021/acs.jctc.8b01196},
	doi = {10.1021/acs.jctc.8b01196},
	number = {3},
	urldate = {2024-02-06},
	journal = {Journal of Chemical Theory and Computation},
	author = {Sayfutyarova, Elvira R. and Hammes-Schiffer, Sharon},
	year = {2019},
	note = {Publisher: American Chemical Society},
	pages = {1679--1689},
}

@article{zhao_casscfcaspt2_2009,
	title = {A {CASSCF}/{CASPT2} study on the low-lying excited states of {HSiCN}, {HSiNC} and their ions},
	volume = {124},
	issn = {1432-2234},
	url = {https://doi.org/10.1007/s00214-009-0585-1},
	doi = {10.1007/s00214-009-0585-1},
	language = {en},
	number = {1},
	urldate = {2024-02-06},
	journal = {Theoretical Chemistry Accounts},
	author = {Zhao, Zeng-Xia and Hou, Chun-Yuan and Shu, Xin and Zhang, Hong-Xing and Sun, Chia-chung},
	year = {2009},
	pages = {85--93},
}

@article{levine_casscf_2020,
	title = {{CASSCF} with {Extremely} {Large} {Active} {Spaces} {Using} the {Adaptive} {Sampling} {Configuration} {Interaction} {Method}},
	volume = {16},
	issn = {1549-9618},
	url = {https://doi.org/10.1021/acs.jctc.9b01255},
	doi = {10.1021/acs.jctc.9b01255},
	number = {4},
	urldate = {2024-02-06},
	journal = {Journal of Chemical Theory and Computation},
	author = {Levine, Daniel S. and Hait, Diptarka and Tubman, Norm M. and Lehtola, Susi and Whaley, K. Birgitta and Head-Gordon, Martin},
	year = {2020},
	note = {Publisher: American Chemical Society},
	pages = {2340--2354},
}

@article{smith_cheap_2017,
	title = {Cheap and {Near} {Exact} {CASSCF} with {Large} {Active} {Spaces}},
	volume = {13},
	issn = {1549-9618},
	url = {https://doi.org/10.1021/acs.jctc.7b00900},
	doi = {10.1021/acs.jctc.7b00900},
	number = {11},
	urldate = {2024-02-06},
	journal = {Journal of Chemical Theory and Computation},
	author = {Smith, James E. T. and Mussard, Bastien and Holmes, Adam A. and Sharma, Sandeep},
	year = {2017},
	note = {Publisher: American Chemical Society},
	pages = {5468--5478},
}

@article{helmich-paris_benchmarks_2019,
	title = {Benchmarks for {Electronically} {Excited} {States} with {CASSCF} {Methods}},
	volume = {15},
	issn = {1549-9618},
	url = {https://doi.org/10.1021/acs.jctc.9b00325},
	doi = {10.1021/acs.jctc.9b00325},
	number = {7},
	urldate = {2024-02-06},
	journal = {Journal of Chemical Theory and Computation},
	author = {Helmich-Paris, Benjamin},
	year = {2019},
	note = {Publisher: American Chemical Society},
	pages = {4170--4179},
}

@article{wouters_density_2014,
	title = {The density matrix renormalization group for ab initio quantum chemistry},
	volume = {68},
	issn = {1434-6079},
	url = {https://doi.org/10.1140/epjd/e2014-50500-1},
	doi = {10.1140/epjd/e2014-50500-1},
	language = {en},
	number = {9},
	urldate = {2024-02-06},
	journal = {The European Physical Journal D},
	author = {Wouters, Sebastian and Van Neck, Dimitri},
	year = {2014},
	pages = {272},
}

@article{marti_density_2010,
	title = {The {Density} {Matrix} {Renormalization} {Group} {Algorithm} in {Quantum} {Chemistry}},
	volume = {224},
	copyright = {De Gruyter expressly reserves the right to use all content for commercial text and data mining within the meaning of Section 44b of the German Copyright Act.},
	issn = {2196-7156},
	url = {https://www.degruyter.com/document/doi/10.1524/zpch.2010.6125/html},
	doi = {10.1524/zpch.2010.6125},
	language = {en},
	number = {3-4},
	urldate = {2024-02-06},
	journal = {Zeitschrift für Physikalische Chemie},
	author = {Marti, Konrad Heinrich and Reiher, Markus},
	year = {2010},
	note = {Publisher: De Gruyter (O)},
	pages = {583--599},
}

@article{sharma_density_2019,
	title = {Density matrix renormalization group pair-density functional theory ({DMRG}-{PDFT}): singlet-triplet gaps in polyacenes and polyacetylenes},
	volume = {10},
	shorttitle = {Density matrix renormalization group pair-density functional theory ({DMRG}-{PDFT})},
	url = {https://pubs.rsc.org/en/content/articlelanding/2019/sc/c8sc03569e},
	doi = {10.1039/C8SC03569E},
	language = {en},
	number = {6},
	urldate = {2024-02-06},
	journal = {Chemical Science},
	author = {Sharma, Prachi and Bernales, Varinia and Knecht, Stefan and Truhlar, Donald G. and Gagliardi, Laura},
	year = {2019},
	note = {Publisher: Royal Society of Chemistry},
	pages = {1716--1723},
}

@article{liebermann_importance-sampling_2022,
	title = {Importance-sampling {FCIQMC}: {Solving} weak sign-problem systems},
	volume = {157},
	issn = {0021-9606},
	shorttitle = {Importance-sampling {FCIQMC}},
	url = {https://doi.org/10.1063/5.0107317},
	doi = {10.1063/5.0107317},
	number = {12},
	urldate = {2024-02-06},
	journal = {The Journal of Chemical Physics},
	author = {Liebermann, Niklas and Ghanem, Khaldoon and Alavi, Ali},
	year = {2022},
	pages = {124111},
}

@article{weser_chemical_2021,
	title = {Chemical insights into the electronic structure of {Fe}({II}) porphyrin using {FCIQMC}, {DMRG}, and generalized active spaces},
	volume = {121},
	copyright = {© 2020 The Authors. International Journal of Quantum Chemistry published by Wiley Periodicals LLC.},
	issn = {1097-461X},
	url = {https://onlinelibrary.wiley.com/doi/abs/10.1002/qua.26454},
	doi = {10.1002/qua.26454},
	language = {en},
	number = {3},
	urldate = {2024-02-06},
	journal = {International Journal of Quantum Chemistry},
	author = {Weser, Oskar and Freitag, Leon and Guther, Kai and Alavi, Ali and Li Manni, Giovanni},
	year = {2021},
	pages = {e26454},
}

@book{shavitt_many-body_2009,
	location = {Cambridge},
	title = {Many-Body Methods in Chemistry and Physics: {MBPT} and Coupled-Cluster Theory},
	isbn = {978-0-521-81832-2},
	url = {https://www.cambridge.org/core/books/manybody-methods-in-chemistry-and-physics/D12027E4DAF75CE8214671D842C6B80C},
	series = {Cambridge Molecular Science},
	shorttitle = {Many-Body Methods in Chemistry and Physics},
	publisher = {Cambridge University Press},
	author = {Shavitt, Isaiah and Bartlett, Rodney J.},
	year = {2009},
	urldate = {2023-07-03},
	date = {2009},
	doi = {10.1017/CBO9780511596834},
}

\end{document}